\documentclass[letterpaper,twocolumn,10pt]{article}
\usepackage[hyphens]{url}      
\usepackage{usenix2019_v3}
\usepackage{tikz}
\usepackage{amsmath}
\usepackage{filecontents}
\usepackage{balance}  
\usepackage{graphics} 
\usepackage[T1]{fontenc}
\usepackage{txfonts}
\usepackage{mathptmx}
\usepackage{color}
\usepackage{booktabs}
\usepackage{textcomp}
\usepackage{fancybox}
\usepackage{microtype} 
\usepackage{ccicons}  
\usepackage{subcaption}
\usepackage{multirow} 
\usepackage{xcolor}
\usepackage{colortbl}
\usepackage{ntheorem}
\usepackage{adjustbox}
\usepackage{array}
\usepackage{booktabs}
\usepackage{multirow}
\usepackage{hhline}
\usepackage{etoolbox} 

\usepackage{todonotes}
\usepackage{listings}
\usepackage{xspace}
\usepackage{cite}
\usepackage{thmtools}
\usepackage{dcolumn}
\usepackage[flushleft]{threeparttable}
\usepackage{enumitem}
\usepackage{comment}
\usepackage[normalem]{ulem}
\usepackage{flushend}
\usepackage{enumitem}

\excludecomment{confidential}

\def\plaintitle{\Large \bf An Observational Investigation of Reverse Engineers' Processes}
\def\plainauthor{
{\rm Daniel Votipka, Seth M. Rabin, Kristopher Micinski*,}\\
{\rm Jeffrey S. Foster$^\dagger$, and Michelle M. Mazurek}\\
       University of Maryland; *Syracuse University; $^\dagger$Tufts University\\
       \{dvotipka,srabin,mmazurek\}@cs.umd.edu; kkmicins@syr.edu; jfoster@cs.tufts.edu
}

\usepackage{cite}

\newcolumntype{d}[1]{D{.}{.}{#1}}
\newcommand{\comm}[3][\color{red}]{{#1{[{#2}: {#3}]}}}
\newcommand{\deletetext}[1]{}

\newcommand{\dan}[1]{\comm[\color{orange}]{Dan}{#1}}
\newcommand{\michelle}[1]{\comm[\color{cyan}]{Michelle}{#1}}

\renewcommand{\paragraph}[1]{\vspace{-.2cm} \hfill \break \textbf{#1.}}

\newcommand{\pone}{P01M\xspace}
\newcommand{\ptwo}{P02V\xspace}
\newcommand{\pthree}{P03V\xspace}
\newcommand{\pfour}{P04V\xspace}
\newcommand{\pfive}{P05V\xspace}
\newcommand{\psix}{P06V\xspace}
\newcommand{\pseven}{P07V\xspace}
\newcommand{\peight}{P08V\xspace}
\newcommand{\pnine}{P09V\xspace}
\newcommand{\pten}{P10B\xspace}
\newcommand{\peleven}{P11M\xspace}
\newcommand{\ptwelve}{P12V\xspace}
\newcommand{\pthirteen}{P13V\xspace}
\newcommand{\pfourteen}{P14M\xspace}
\newcommand{\pfifteen}{P15V\xspace}
\newcommand{\psixteen}{P16M\xspace}
\newcommand{\Overview}{Overview}
\newcommand{\Skimming}{Sub-component scanning}
\newcommand{\Indepth}{Focused experimentation}
\newcommand{\overview}{overview}
\newcommand{\skimming}{sub-component scanning}
\newcommand{\indepth}{focused experimentation}
\newcommand{\OVERVIEW}{Overview}
\newcommand{\SKIMMING}{Sub-component Scanning}
\newcommand{\INDEPTH}{Focused Experimentation}
\newcommand{\Gone}{Match interaction with analysis phases}
\newcommand{\Gtwo}{Present input and output in the context of code}
\newcommand{\Gthree}{Allow data transfer between static and dynamic contexts}
\newcommand{\Gfour}{Allow selection of analysis methods}
\newcommand{\Gfive}{Support readability improvements}
\newcommand{\beacon}{beacon}

\newcommand{\beacons}{beacons}
\newcommand{\Beacons}{Beacons}
\newcommand{\RE}{RE}
\newcommand{\REs}{REs}

\newcolumntype{L}[1]{>{\raggedright\let\newline\\\arraybackslash\hspace{0pt}}m{#1}}
\newcolumntype{C}[1]{>{\centering\let\newline\\\arraybackslash\hspace{0pt}}m{#1}}
\newcolumntype{R}[1]{>{\raggedleft\let\newline\\\arraybackslash\hspace{0pt}}m{#1}}

\definecolor{dkgreen}{rgb}{0,0.6,0}
\definecolor{gray}{rgb}{0.5,0.5,0.5}
\definecolor{mauve}{rgb}{0.58,0,0.82}

\begin{document}

\date{}
%
\title{\plaintitle}

\author{\plainauthor}


%


\maketitle

\begin{abstract}
Reverse engineering is a complex process essential to software-security tasks such as  
vulnerability discovery and malware analysis. Significant research and engineering effort 
has gone into developing tools to support reverse engineers. However, little work has been 
done to understand the way reverse engineers think when analyzing programs, leaving tool 
developers to make interface design decisions based only on  intuition.

This paper takes a first step toward a better understanding of reverse engineers' processes, with the goal of producing insights for improving interaction design for reverse 
engineering tools. We present the results of a semi-structured, observational interview study 
of reverse engineers (N=16). Each observation investigated the questions reverse engineers 
ask as they probe a program, how they answer these questions, and the decisions 
they make throughout the reverse engineering process. From the interview responses, 
we distill a model of the reverse engineering process, divided into 
three phases: \overview{}, \skimming{}, and \indepth{}. Each analysis phase's results 
feed the next as reverse engineers' mental representations become more concrete. We find 
that reverse engineers typically use static methods in the first two phases, but dynamic
methods in the final phase, with experience playing large, but varying, roles in each phase. 
Based on these 
results, we provide five interaction design guidelines for reverse engineering tools.
\end{abstract}


%

\section{Introduction}
\label{sec:intro}
Software reverse engineering is a key task performed by security professionals during vulnerability discovery, malware analysis, and other tasks~\cite{Votipka2018HackersTesters,HackersReadObfuscatedCeccato2017},~\cite[pg. 5-7]{eilam2011reversing}. 
(For brevity, we will refer to this task as \RE{} and its practitioners as \REs{}.) 
\RE{} can be complex and time consuming, often requiring expert knowledge and 
extensive experience to be successful~\cite{chess,Yakan2016DREAM++}. In one study, participants analyzing 
small decompiled code snippets with less than 150 lines required 39 minutes on average to answer 
common malware-analysis questions~\cite{Yakan2016DREAM++}.


Researchers, companies, and practitioners have developed an extensive array of tools to support \RE{}~\cite{ShoshitaishviliHaCRS2017, Rutar2004, BacaCPL13, Doupe2010, Austin2011, Antunes2009, suto2007, suto2010, McGraw2011,Yakan2016DREAM++,Enck2010, cadar2008klee, Cha2012,stephens2016driller,ida,binaryninja,coverity,forallsecure, idaplugins,binaryninjaplugins}. 
However, there is limited theoretical 
understanding of the \RE{} process itself. While existing tools are quite useful, design decisions are currently ad-hoc and based on each designer's personal experience. With a more rigorous
and structured theory of \REs{}' processes, habits, and mental models, we believe existing tools could 
be refined, and even better tools could be developed.
\begin{confidential}
\deletetext{, we believe they could be 
refined---and even better tools could be developed---with a better understanding of \REs{}' processes, 
habits, and mental models.} 
\end{confidential}
This follows from recommended design principles for tools supporting complex, exploratory tasks, 
in which the designer should ``pursue the goal of having the computer vanish"~\cite[pg. 19-22]{Shneiderman2016DUI}. 

In contrast to \RE{}, there is significant theoretical understanding of more traditional program 
comprehension---how developers read and understand program functionality---including tasks such as program maintenance and debugging~\cite{Letovsky1986,PCFactFindingLaToza2007,CognitivePCArunachalam1996,ProDevPCRoehm2012,Gugerty1986,Brooks1983,vonMayrhauser1995,SchemaBasedDetienne1990,DevsSeekKo2006,StimulusStructuresPennington1987,Littman1986}. However, \RE{} differs from these tasks, as \REs{} typically do not have access to the original source, the developers who wrote the program, or internal documentation~\cite[pg. 141-196]{eilam2011reversing},~\cite{Chikofsky1990}. Further, \REs{} often must overcome countermeasures, such as symbol stripping, packing, obfuscation, and anti-debugging techniques~\cite[pg. 327-356]{eilam2011reversing},~\cite{OKane2011},~\cite[pg. 441-481]{ligh2010malware},~\cite[pg. 660-661]{harper2018gray}.
As a result, it is unclear which aspects of traditional program comprehension processes will translate 
to \RE{}. 

In this paper, we develop a theoretical model of the \RE{} process, 
with an eye toward building more intuitive \RE{} tools. 
In particular, we set out to answer the following research questions:

\begin{enumerate}[label=\textbf{RQ\arabic*.}, leftmargin=*, itemsep=0.5ex]
	\item What high-level process do \REs{} follow when examining a new program?
	\item What technical approaches (i.e., manual and automated analyses) do \REs{} use?
	\item How does the \RE{} process align with traditional program comprehension? How does it differ?
\end{enumerate}

Specifically, when considering \REs{}' processes, we sought to determine the types of questions they had to answer and hypotheses they generated; the specific steps taken to learn more about the program; and the way they make decisions throughout the process (e.g., which code segments to investigate or which analyses to use).

As there is limited prior work outlining \REs{}' processes and no theoretical basis on which to build quantitative assessments, we chose an exploratory qualitative approach, building on prior work in expert decision-making~\cite{klein1989recognition,klein1986rapid,cannon1998making} and program comprehension~\cite{Letovsky1986,PCFactFindingLaToza2007,CognitivePCArunachalam1996,ProDevPCRoehm2012,Gugerty1986,Brooks1983,vonMayrhauser1995,SchemaBasedDetienne1990,DevsSeekKo2006,StimulusStructuresPennington1987,Littman1986}. While a qualitative study cannot indicate prevalence or effectiveness of any particular process, it does allow us to enumerate the range of \RE{} behaviors and investigate in depth their characteristics and interactions. Through this study, we can create a theoretical model of the \RE{} process as a reference for future tool design.

To this end, we conducted a 16-participant, semi-structured observational study. In each participant session, we asked participants to recreate a recent \RE{} experience while we observed their actions and probed their thought process. Throughout, we tracked the decisions made, mental simulation methods used, questions asked, hypotheses formulated, and \beacons{} (recognizable patterns) identified. 

We found that in general, the \RE{} process can be modeled in three phases: \overview{}, \skimming{}, and \indepth{}. 
\begin{confidential}
\deletetext{Based on a high-level goal, such as finding a vulnerability or identifying malicious behaviors,}
\end{confidential} 
\REs{} begin by establishing a broad view of the program's functionality (\textit{\overview{}}). They use their overview's results to prioritize sub-components---e.g., functions---for further analysis, only performing detailed review of specific sub-components deemed most likely to yield useful results (\textit{\skimming{}}). As \REs{} review these sub-components, they identify hypotheses and questions that are tested and answered, respectively, through execution or in-depth, typically manual static analysis (\textit{\indepth{}}). The last two phases form a loop. \REs{} develop hypotheses and questions, address them, and use the results to inform their understanding of the program. This produces new questions and hypotheses, and the \RE{} continues to iterate until the overall goal is achieved. 

Further, we identified several trends in \REs{}' processes spanning multiple phases. We found that \REs{} use more static analysis in the first two phases and switch to dynamic simulation methods during \indepth{}. We also observed that experience plays an important role throughout \REs{}' decision-making processes, helping \REs{} prioritize where to search (\overview{} and \skimming{}), recognize implemented functionality and potential vulnerabilities (\skimming{}), and select which mental simulation method to employ (all phases). Finally, we found \REs{} choose to use tools to support their analysis when a tool's input and output can be closely associated with the code and when the tools improve code readability.

Based on these results, we suggest five guidelines for designing \RE{} tools.

\section{Background and Related Work}
\label{sec:background}
While little work has investigated expert \RE{}, there has been significant effort studying 
similar problems of naturalistic decision-making (NDM) and program comprehension. Because of their similarity, 
we draw on theory and methods that have been found useful in these areas~\cite{klein1989critical,klein2017sources,Letovsky1986,PCFactFindingLaToza2007,CognitivePCArunachalam1996,ProDevPCRoehm2012,Gugerty1986,Brooks1983,vonMayrhauser1995} as well as in initial studies of \RE{}~\cite{Bryant2012}. 

\subsection{Naturalistic Decision-Making}
\label{sec:ndm}
Significant prior work has investigated how experts make decisions in real-world (naturalistic) situations and the factors that influence them. Klein et al. proposed the theory of Recognition-Primed Decision-Making (RPDM)~\cite[pg. 15-33]{klein2017sources}. The RPDM model suggests experts recognize components of the current situation---in our case, the program under investigation---and quickly make judgments about the current situation based on experiences from prior, similar situations. Therefore, experts can quickly leverage prior experience to solve new but similar problems. Klein et al.\ have shown this decision-making model is used by firefighters~\cite{klein1989recognition,klein1986rapid}, military officers~\cite{ross2004recognition,cannon1998making}, medical professionals~\cite[pg. 58-68]{zsambok2014naturalistic}, and software developers~\cite{klein1986}. Votipka et al. found that vulnerability-discovery experts rely heavily on prior experience~\cite{Votipka2018HackersTesters}, suggesting that RPDM may be the decision-making model they use. 

%
NDM research focuses on these decision-making processes and uses interview techniques designed to highlight critical decisions, namely the Critical Decision Method, which has participants walk through specific notable experiences while the interviewer records and asks probing follow-up question about items of interest to the research (see Section~\ref{sec:protocol})~\cite{klein1989critical}. Using this approach prior work has driven improvements in automation design. Specifically, these methods have identified tasks within expert processes for automation~\cite{klein1989critical,Klein1997}, and inferred mental models used to support 
effective interaction design~\cite{Rasmussen1983} in several domains, including automobile safety controls~\cite{Yamaguchi2000,Ohno2000}, military decision support~\cite{Worm2000,Akbari2000,klein1989critical,miller1992decision}, and manufacturing~\cite{Ohtsuka1997,Klinger1992}. 
\begin{confidential}
\dan{Possibly cut the specific references in the sentence above and just put all the cites after ``several domains''.}
\end{confidential}
Building on its demonstrated success, we apply the 
Critical Decision Method to guide our investigation.


\subsection{Program Comprehension}
\label{sec:pc}
Program comprehension research investigates how developers maintain, 
modify, and debug unfamiliar code---similar problems to \RE{}. Researchers have found that developers approach unfamiliar programs from a non-linear, fact-finding 
perspective~\cite{Letovsky1986,PCFactFindingLaToza2007,CognitivePCArunachalam1996,ProDevPCRoehm2012,Gugerty1986,Brooks1983,vonMayrhauser1995}. They 
make hypotheses about program functionality and focus on proving or disproving their hypotheses. 

Programmers' hypotheses are based on \textit{\beacons{}} recognized when scanning through the program. \Beacons{} are common schemas or patterns, which inform how developers expect variables and program components to behave~\cite{SchemaBasedDetienne1990,CognitivePCArunachalam1996,DevsSeekKo2006,StimulusStructuresPennington1987}. To evaluate their hypotheses, developers either mentally simulate the program by reading it line 
by line, execute it using targeted test cases, or search for other \beacons{} that contradict their 
hypotheses~\cite{SchemaBasedDetienne1990,HackersReadObfuscatedCeccato2017,CognitivePCArunachalam1996,ProDevPCRoehm2012,Littman1986}. Von Mayrhauser and Lang showed developers switch among these methods regularly, depending on the program context or hypothesis~\cite{vonMayrhauser1998program}.  Further, when reading code, developers focus on data- and control-flow 
dependencies to and from their \beacons{} of interest~\cite{DevsSeekKo2006,LaToza2010DAR}.

We anticipated that \REs{} might exhibit similar behaviors, so we build on this prior work by focusing on hypotheses, \beacons{}, and simulation methods during interviews (Section~\ref{sec:protocol}). 
However, we also hypothesized some process divergence, as \RE{} and ``standard'' program comprehension 
differ in several key respects. Reverse engineers generally operate on obfuscated code and raw binaries, which are harder to 
read than source code. Further, \REs{} often focus on identifying and exploiting flaws in the program, 
instead of adding new functionality or fixing known errors. 
\vspace{-.17cm}
\subsection{Improving Usability for \RE{} Tools}
Several researchers have taken steps to improve \RE{} tool usability. Do et al.\ created a Just-in-time static analysis framework called CHEETAH, based on the result of user studies investigating how developers interact with static analysis tools~\cite{Johnson2013,Smith2015StaticAnalysisQuestions}. CHEETAH lets developers run static analyses incrementally as they write new code, allowing developers to put the analyses results in context and reduce the overwhelming ``wall of alerts'' feeling. While we follow a similar qualitative approach, we focus on a different population (i.e., \REs{} instead of developers) and task (\RE{} instead of security alert response).

Shoshitaishvili et al. propose a tool-centered human-assisted vulnerability discovery paradigm~\cite{ShoshitaishviliHaCRS2017}. They suggest a new interaction pattern where users provide on-demand feedback to a automated agent by performing well-defined sub-tasks to support the agent's analysis. This model leverages human insights to overcome the automation's deficiencies, outperforming the best automated systems while allowing the analysis to scale significantly beyond limited human resources. However, the demonstrated interaction model specifically targets non-expert users who do not understand program internals (e.g., code, control flow diagrams, etc.), treating the program as a black box.

Focusing on expert users, Kruger et al. propose a specification language to allow cryptography experts to state secure usage requirements for cryptographic APIs~\cite{krgeretal:LIPIcs:2018:9215}. Unfortunately, this approach still requires the expert to learn a new, potentially complicated language---hundreds of lines of code for each API.

Finally, Yakdan et al.\ designed a decompiler, DREAM++, intended to improve usability compared to existing 
tools~\cite{Yakan2016DREAM++}. 
DREAM++'s experimental evaluation showed that a simple set of code transformations significantly increased both students' and professionals' ability to \RE{} malware, demonstrating the benefit of even minor usability improvements.
We hope that our more complete investigation of \REs{}' processes may spur the development of further high-impact improvements.


\subsection{The Vulnerability Discovery Process}
Ceccato et al.\ reviewed detailed reports by three penetration testing teams searching for vulnerabilities in a suite of security-specific programs~\cite{HackersReadObfuscatedCeccato2017}. The participating teams were asked to record their process for searching the programs, finding vulnerabilities, and exploiting them. Our study delves deeper into the specific problem of \RE{} a program to understand its functionality. Further, through our interviews, we are able to probe the \RE{}'s process to elicit more detailed responses.

Most similarly to this work, Bryant investigated \RE{} using a mixed methods approach, including three semi-structured 
interviews with \REs{} and an observational study where four participants completed a predesigned \RE{} task~\cite{Bryant2012}. Based on his observations, Bryant developed a sense-making model for reverse engineering where \REs{} generate hypotheses from prior experience and cyclically attempt to (in)validate these hypotheses, generating new hypotheses in the process. Our results align with these findings; we expand on them, producing a more detailed model describing the specific approaches used and how \RE{} behaviors change throughout the process. Our more detailed model is achieved through our larger sample size and observation of \RE{} processes on different, real-world programs, demonstrating \RE{} behaviors to ensure saturation of themes~\cite[pg. 113-115]{charmazconstructing}.

In our prior work, we performed 25 interviews of white-hat hackers and testers to determine their vulnerability discovery processes \cite{Votipka2018HackersTesters}. While this research identified \RE{} as an important part of the vulnerability discovery process, its broader focus (e.g., process, skill development, and community interaction) limited its ability to provide details regarding how \RE{} is carried out, leading us to our current, more focused investigation.

\section{Method}
\label{sec:method}

We are interested in developing a theoretical model of the \RE{} process with respect to both overall strategy and specific techniques used. In particular, we focus on the three research questions given in Section~\ref{sec:intro}.
\begin{confidential}
\deletetext{In particular, we pose the following research questions:}

\begin{enumerate}[label=\textbf{\deletetext{RQ\arabic*.}}, leftmargin=*]
	\item \deletetext{What high-level process do \REs{} follow when examining a new program?}
	\item \deletetext{ What technical approaches (i.e., manual and automated analyses) do \REs{} use?}
	\item \deletetext{ How does the \RE{} process align with traditional program comprehension? How does it differ?}
\end{enumerate}
\end{confidential}

\begin{confidential}
\deletetext{Answers to these questions can support the development of more intuitive \RE{} tools and provide guidance about how to modify existing tools to improve usability.}
\end{confidential}
To answer these questions, we employ a semi-structured, observation-based interview protocol, designed to yield detailed insights into \RE{} experts' processes. The full protocol is given in Appendix~\ref{appendix:protocol}. Interviews lasted 70 minutes on average. Audio and video were recorded during each interview. All interviews were led by the first author, who has six years of professional RE experience, allowing him to understand each RE's terminology and process, ask appropriate probing questions, and identify categories of similar actions for coding. Participants were provided a \$40 gift card in appreciation of their time. Our study was reviewed and approved by the University of Maryland's Institutional Review Board. In this section, we describe our 
\begin{confidential}
\deletetext{recruitment methods, }
\end{confidential}
interview protocol and data analysis process, and we discuss limitations of our method. 
%

\subsection{Interview Protocol}
\label{sec:protocol}
We performed semi-structured, observational video-teleconference interviews. We implemented a modified version of the Critical Decision Method, which is intended to reveal expert knowledge by inquiring about specific cases of interest~\cite{klein1989critical}. We asked participants to choose an interesting program they recently reverse engineered, and had them recall and demonstrate the process they used. Each observation was divided into the two parts: 
program background and \RE{} process. Throughout, the interviewer noted and asked further questions about multiple items of interest.

\paragraph{Program background} 
We began by asking participants to describe the program they chose to reverse engineer. This included questions about the program's functionality and size, what tools (if any) they used, and whether they reverse engineered the program with others. 

\paragraph{Reverse engineering process}
Next, we asked participants about their program-specific \RE{} goals, and then asked them to recreate their process while sharing their screen (RQ1)\footnote{The only participant who did not share their screen did so because of technical difficulties that could not be resolved in a timely manner.}. We chose to have participants demonstrate their process, asking them to open all tools they used and perform all original steps, so we could observe automatic and subconscious behaviors---common in expert tasks~\cite{annett2003hierarchical}---that might be missed if simply asked to recall their process. As the participant recreated their process, we asked several directed questions intended to probe their understanding while allowing them to delve into areas they felt were important.
We encouraged participants to share their entire process, even if a particular speculative step did not end up  
supporting their final goal. For example, they may have decided to reverse a function that turned out to be a common library function already documented elsewhere, resulting in no new information gain.

Instead of asking participants to demonstrate a recent experience, we could have asked them to \RE{} a program new to them. This could be more representative of the real-world experience of approaching a new program and might highlight additional subconscious or automatic behaviors. However, it would likely require a much longer, probably unreasonable period of observation. When asked how much time participants spent reverse engineering the programs demonstrated, answers ranged from several hours to weeks. Alternatively, we could have asked participants to \RE{} a toy program. However, this approach restricts the results, both in depth of process and in terms of the program type(s) selected. Demonstration provides a reasonable compromise, and is a standard practice in NDM studies~\cite{klein1989critical}. In practice, we believe the effect of demonstration was small, especially because the interviewer asked probing questions to reveal subconscious actions. 


\paragraph{Items of interest}
The second characteristic of the Critical Decision Method is that the interviewer asks follow-on questions about items of interest to the research. We selected our items of interest from those identified as important in prior NDM (\textit{decision}) and program comprehension (\textit{questions/hypotheses}, \textit{\beacons{}}, \textit{simulation methods}) literature---discussed in Sections~\ref{sec:ndm} and~\ref{sec:pc}, respectively. These items were chosen to identify specific approaches used (RQ2) and differences between \RE{} and other program comprehension tasks (RQ3). Below, we provide a short description of each and a summary of follow-on questions asked:

$\bullet$ \textbf{Decisions.} These are moments where the \RE{} decides between one or more actions. This can include deciding whether to delve deeper into a specific function or which simulation method to apply to validate a new hypothesis. For decision points, we asked participants to explain how they made the decision. For example, when deciding to analyze a function, the \RE{} might consider what data flows into the function as arguments or what calls it.

$\bullet$ \textbf{Questions/Hypotheses.} These are questions that must be answered or conjectures about what the program does. Reverse engineers might form a hypothesis about the main purpose of a function, or whether a certain control flow is possible. Prior work has shown that hypotheses are central part to program comprehension~\cite{PCFactFindingLaToza2007,CognitivePCArunachalam1996,ProDevPCRoehm2012,HackersReadObfuscatedCeccato2017}, so we expected hypothesis generation and testing to be central to \RE{}. For hypotheses, we asked participants to explain why they think the hypothesis might be true and how they tested it. As an example, if a \RE{} observes a call to \lstinline{strcpy}, they might hypothesize that a buffer overflow is possible. To validate their hypothesis, they would check whether unbounded user input can reach this call.

$\bullet$ \textbf{Simulation methods.} Any process where a participant reads or runs the code to determine its function.
We asked \REs{} about any manual or automated simulation methods used: for example, using a debugger to determine the program's memory state at a specific point. We wanted to know whether they employed any tools and if they were custom, open source, or purchased. Further, we asked them to evaluate any tools used, and to discuss their effectiveness for this particular task. 
\begin{confidential}
\dan{I don't think we really talk much about the tool types/decisions in the results, so it might not make sense to discuss it here.}
\end{confidential}
Additionally, we asked participants why they used particular simulation methods, whether they typically did so, the method's inputs and outputs, and how they know when to switch methods.

$\bullet$ \textbf{\Beacons{}.} These include patterns or tells that a \RE{} recognizes, allowing them to quickly generate hypotheses about the program's functionality without reading line-by-line. For example, if a \RE{} sees an API call to get a secure random number with several bit-shift operations, they may assume the associated function performs a cryptographic process.
For \beacons{}, we had \REs{} explain why the \beacon{} stood out and how they recognized it as that sort of \beacon{} rather than some other pattern. The goal in inquiring into this phenomenon is to understand how \REs{} perform pattern matching, and identify potentially common \beacons{} of importance.

Additionally, we noted whenever participants referenced documentation or information sources external to the code---e.g., StackOverflow, \RE{} blogs, API documentation---to answer a program functionality question. We asked whether they use that resource often, and why they selected that resource.

To make the interviews more fluid and less repetitive, we intentionally skipped questions that had already been answered in response to prior questions. To ensure consistency, all the interviews were conducted by the first author.

We conducted two pilot interviews prior to the main study. After the first pilot, we made adjustments to ensure appropriate terminology was used and improve question flow. However, no changes were required after the second interview, so we included the second pilot interview in our main study data.

\subsection{Data Analysis}
We applied iterative open coding to identify interview themes~\cite[pg. 101-122]{strauss1990basics}. After completing each interview, the audio was sent to an external transcription service. 
The interviewer and another researcher first collaboratively coded three interviews---reviewing both the text and video---to create an initial codebook.
Then, the two coders independently coded 13 interviews, comparing codes after every three interviews to determine inter-coder reliability. To measure inter-coder reliability, we used Krippendorff's Alpha ($\alpha$), as it accounts for chance agreements~\cite{kripalpha}.\footnote{The ReCal2 software package was used to calculate Krippendorff's Alpha~\cite{freelon2010recal}} After each round, the coders resolved any differences, updated the codebook as necessary, and re-coded previously coded interviews. The coders repeated this process four times until they achieved an $\alpha$ of 0.8, which is above the recommended level for exploratory studies~\cite{kripalpha,lombard2002content}. 
The final codebook is given in Appendix~\ref{appendix:book}.

Next, we sought to develop our theoretical model by extracting themes from the coded data. 
First, we grouped identified codes into related categories. Specifically, we discovered three categories associated with the phases of analyses performed by \REs{} (i.e., \OVERVIEW{}, \SKIMMING{}, and \INDEPTH{}). Then, we performed an axial coding to determine relationships between and within each phase and trends across the three phases~\cite[pg. 123-142]{strauss1990basics}. From these phases and their connections, we derive a theory of \REs{}' high-level processes and specific technical approaches. We also present a set of interaction-design guidelines for building analysis tools to best fit \REs{}.
\begin{confidential}
\dan{We probably don't need the last two sentences.  They are kind of just hitting the same points again.}
\end{confidential}

\subsection{Limitations}
There are a number of limitations innate to our methodology. First, participants likely do not recall all task details they are asked to relay. This is especially common for expert tasks~\cite{annett2003hierarchical}. We attempt to address this by using the CDM protocol, which has been used successfully in prior decision-making research on expert tasks~\cite{klein1989critical}. Furthermore, we asked participants to recreate the \RE{} task while the interviewer observed. This allowed the interviewer to probe subconscious actions that would likely have been skipped without observation.

Participants also may have skipped portions of their process to protect trade secrets; however, in practice we believe this did not impact our results. Multiple participants stated they could not demonstrate certain confidential steps, but the secret component was in the process's operationalization (e.g., the keyword list used or specific analysis heuristics). In all cases, participants still described their general process, which we were able to include in our analysis.

Finally, we focus on experienced \REs{} to understand and model expert processes. Future work should consider newer \REs{} to understand their struggles and support their development.



\section{Recruitment and Participants}
\label{sec:recandpart}
We recruited interview participants from online forums, vulnerability discovery organizations, and relevant conferences.

\begin{confidential}
\dan{There is a lot of discussion about why we did each recruitment method in here...maybe we don't need that?}
\end{confidential}

\paragraph{Online forums}
We posted recruitment notices on a number of \RE{} forums, including forums for popular \RE{} tools such as IDAPro and BinaryNinja. We also posted ads on online communities like Reddit. Dietrich et al. showed online chatrooms and forums are useful for recruiting security professionals, since participants are reached in a more natural setting where they are more likely to be receptive~\cite{Dietrich2018SecMisconfig}.

\paragraph{Related organizations}
We contacted the leadership of ranked CTF teams\footnote{Found via \url{https://ctftime.org/}} and bug bounty-as-a-service companies asking them to share study details with their members. 
Our goal in partnering with these organizations was to gain credibility with members and avoid our messages dismissed as spam. Prior work found relative success with this strategy~\cite{Votipka2018HackersTesters}. 
To lend further credibility, all emails were sent from an address associated with our institution, and detailed study information was hosted on a web domain owned by our institution.

\paragraph{Relevant conferences}
Finally, we recruited at several conferences commonly attended by \REs{}. We explained study details and participant requirements in person and distributed business cards with study information. Recruiting face-to-face allowed us to clearly explain the goal of the research and its potential benefits to the \RE{} community.

\paragraph{Participant screening}
We asked respondents to our recruitment efforts to complete a short screening questionnaire. Our questionnaire
(see Appendix~\ref{appendix:survey} for full questionnaire)
asked participants to self-report their level of \RE{} expertise on a five-point Likert-scale from novice to expert; indicate their years of \RE{} experience; and answer demographic questions. As our goal is to produce interaction guidelines to fit \REs{}' processes, building on less experienced \REs{}' approaches may not be beneficial. Therefore, we only selected participants who rated themselves at least a three on the Likert scale and had at least three years of \RE{} experience.
We contacted volunteers in groups of ten in random order, waiting one week for their response before moving to the next group. This process continued until we reached sufficient interview participation. 

\paragraph{Participants}
We conducted interviews between October 2018 and January 2019. We received 68 screening survey responses; 42 met our expertise criteria. Of these volunteers, 16 responded to randomly ordered scheduling requests and were interviewed. We stopped further recruitment after 16 interviews, when we reached \emph{saturation}, meaning we no longer observed new themes emerging. This is the standard stopping criteria for a rigorous qualitative process~\cite[pg. 113-115]{charmazconstructing}. Because our participant count is within the range recommended by best practice literature (12-20 participants), our results provide useful insights for later quantitative inquiry and generalizable recommendations~\cite{guest2006many}.

Table~\ref{tab:participants} shows the type of program each participant reverse engineered during the interview and their demographics, including their self-reported skill level, years of experience, and the method used to recruit them. Each participants' ID indicates their assigned ID number and the primary type of \RE{} tasks they perform. For example, \pone{} indicates the first interviewee is a malware analyst. Note that three interviewees used a challenge binary\footnote{An exercise program designed to expose \REs{} to interesting concepts in a simple setting} during the interview. These participants could not show us any examples from their normal work due to the proprietary or confidential nature of their work. Instead, we asked them to discuss where their normal process on a larger program differed from process they showed with the challenge binary.

While we know of no good \RE{} demographics surveys, our participant demographics are similar to bug-bounty hunters, who commonly perform \RE{} tasks. Our population is mostly male (94\%), young (63\% $<30$) and well educated (75\% with a bachelor's degree). HackerOne~\cite{hackeronehackerreport} and Bugcrowd report similar genders (91\% of Bugcrowd hunters), ages (84\% $<35$ and 77\% $<30$, respectively), and education levels (68\% and 63\% with a bachelor's, respectively) for bug-bounty hunters.


\begin{table}[t]
\center
\begin{threeparttable}
\footnotesize
\begin{tabular}{l l l c c l}
        \textbf{ID}\tnote{1} & \textbf{Program} & \textbf{Edu.} & \textbf{Skill}\tnote{2} & \textbf{Exp.} & \textbf{Recruitment} \\
     \toprule
    \pone & Malware & B.S. & 4 & 7 & Conference \\
    \ptwo & System & HS & 4 & 8 & Conference \\
    \pthree & Challenge & M.S. & 4 & 6 & Conference \\
    \pfour & Challenge & B.S. & 5 & 11 & Conference \\
    \pfive & Application & M.S. & 5 & 6 & Forum \\
    \psix & Challenge & HS & 4 & 10 & Forum \\
    \pseven & System & M.S. & 5 & 10 & Forum \\
    \peight & Firmware & Assoc. & 4 & 5 & Forum \\
    \pnine & Firmware & B.S. & 4 & 14 & Forum \\
    \pten & Malware & M.S. & 5 & 15 & Organization \\
    \peleven & Malware & Ph.D. & 3 & 10 & Forum \\
    \ptwelve & System & B.S. & 3 & 8 & Forum \\
    \pthirteen& Application & B.S. & 5 & 21 & Forum \\ 
    \pfourteen & Malware & M.S. & 4 & 5 & Forum \\
    \pfifteen & Application & HS & 3 & 4 & Forum \\
    \psixteen & Malware & M.S. & 3 & 3 & Forum \\
    \bottomrule
\end{tabular}
   \begin{tablenotes}
     \small
      \item[1] M: Malware analysis, V: Vulnerability discovery, B: Both
      \item[2] Scale from 0-5, with 0 indicating no skill and 5 indicating an expert
   \end{tablenotes}
\end{threeparttable}
\caption{Participant demographics.}
\label{tab:participants}
    \vspace{-.4cm}
\end{table}

\section{Results: An \RE{} Process Model}
\label{sec:tiers}
\begin{figure*}
    \centering
    \includegraphics[width=.9\textwidth]{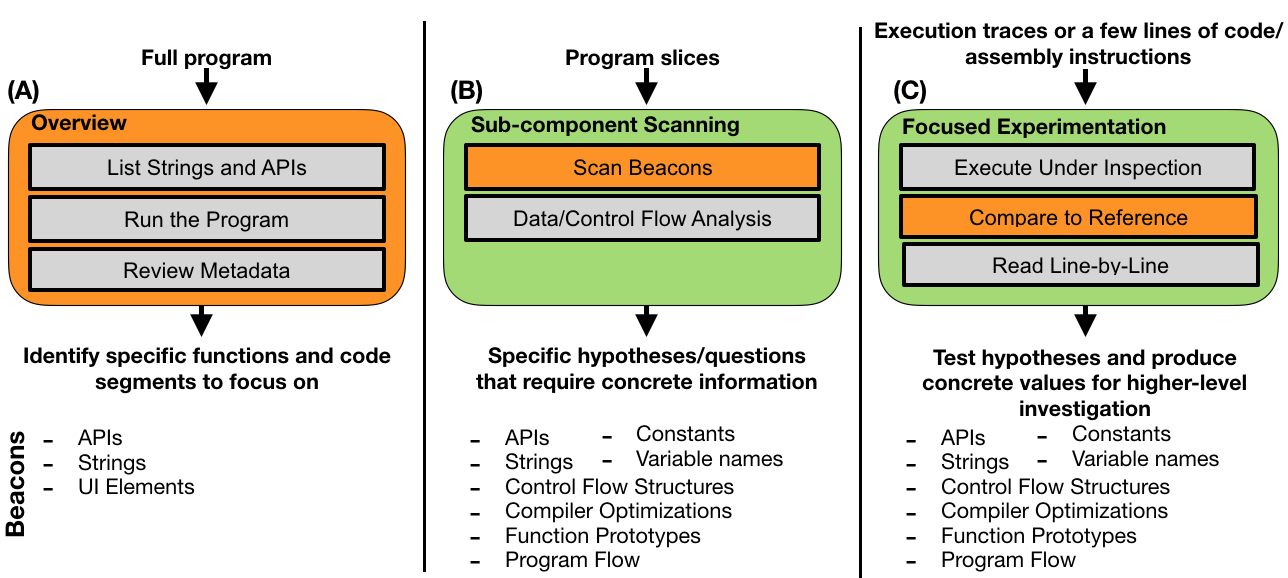}
    \caption{Overview of \REs{}' three analysis phases. For each phase, the analyzed program scope is shown at the top, simulation methods used are in rectangles, and the analysis results are below the phase. Finally, the phase's beacons are at the bottom of the figure. Segments differing the most from the program comprehension literature are colored orange.}
    \label{fig:tiers}
     \vspace{-.5cm}
\end{figure*}

Across all participants, we observed at a high-level (RQ1) their \RE{} process could be divided into three distinct phases: \Overview{}, \Skimming{}, and \Indepth{}.
Beginning with a general goal---e.g., identifying vulnerabilities or malicious behaviors---\REs{} seek a broad overview of the program's functionality (\textit{\overview{}}). They use this to establish initial hypotheses and questions which focus investigation on certain sub-components, in which they only review subsets of information (\textit{\skimming{}}). Their focused review produces more refined hypotheses and questions. Finally, they attempt to test these hypotheses and answer specific questions through execution or in-depth static analysis (\textit{\indepth{}}). Their detailed analysis results are then fed back to the second phase for further investigation, iteratively refining questions and hypotheses until the overall goals are achieved. Each phase has its own set of questions, methods, and \beacons{} that make up the technical approaches taken by \REs{} (RQ2). In this section, we describe each phase in detail and highlight differences between \RE{} and traditional program comprehension tasks (RQ3). In the next section, we discuss trends observed across these phases, including \RE{} process components common to multiple phases, such as factors driving their decision-making. Figure~\ref{fig:tiers} provides an overview of each phase of analysis.

Note, in this section and the next, we give the number of \REs{} who expressed each idea. We include counts to indicate prevalence, but a participant not expressing an idea may only mean they failed to state it, not that they disagree with it. Therefore, we do not perform comparisons between participants using statistical hypothesis tests. It is uncertain whether our results generalize past our sample, but they suggest future work and give novel insights into the human factors of \RE{}.

Somewhat to our surprise, we generally observed the same process and methods used by \REs{} performing both malware analysis and vulnerability discovery. In a sense, malware analysts are also seeking an exploit: a unique execution or code pattern that can be exploited as a signature or used to recover from an attack (e.g.,  ransomware). 
We did observe differences between groups, but only in their operationalization of the analysis process. For example, the two groups focused on different APIs and functionality (e.g., vulnerability finders looked at memory management functions and malware analysts focused on network calls). However, because our focus is on the high-level process and methods used, we discuss both groups together in the following sections.

\begin{confidential}
\michelle{i wonder if we want to add anything here about importance of ideas being expressed by even 1 participant? maybe not.}
\end{confidential}

%
%

\subsection{\OVERVIEW{} (RQ1)}
Reverse engineers may have a short description of the program they are investigating (N=2), some familiarity with its user interface (N=2), or an intuition from prior experience about the functions  the program likely performs (N=7). However, they generally do not have prior knowledge about the program's organization or implementation (N=16). They might guess that the program performs cryptographic functions because it is a secure messaging app, but they do not know the algorithm or libraries used, or where in the code cryptographic protocols are implemented. Therefore, they start by seeking a high-level program view (N=16). This guides which parts of the program to prioritize for more complex investigation. \pone{} said this allows him to ``get more to the core of what is going on with this binary.'' Reverse engineers approach this phase in several ways. The left section of Figure~\ref{fig:tiers} summarizes the \overview{} phase's simulation methods, \beacons{}, and outputs. We discuss these items in more detail below.
\begin{confidential}
\dan{We can probably cut the last sentence}
\end{confidential}

\paragraph{Identify the strings and APIs used (RQ2)} Most \REs{} begin by listing the strings and API calls used by the program (N=15). These lists allow them to quickly identify interesting components. \pthree{} gave the example that ``if this was a piece of malware\ldots{}and I knew that it was opening up a file or a registry entry, I would go to imports and look for library calls that make sense. Like \lstinline{refile} could be a good one. Then I would find where that is called to find where malicious behavior starts.'' In some cases, \REs{} begin with specific functionality they expect the program to perform and search for related strings and APIs (N=7). As an example, \peight{} performed a ``grep over the entire program looking for \lstinline{httpd} because a lot of times these programs have a watchdog that includes a lot of additional configuration details.'' 

\paragraph{Run the program and observe its behavior (RQ2)} Many \REs{} execute the program to see how it behaves under basic usage (N=7).
When running the program, some \REs{} look at UI elements (e.g., error messages), then search for them in the code, marking associated program components for further review (N=3). For example, \pthirteen{} began by ``starting the software and looking for what is being done.'' He was shown a pop-up that said he had limited features with the free version. He observed that there was ``no place I can put a [access] code, so it must be making a web services check'' to determine license status. Next, he opened the program in a disassembler and searched for the pop-up's text ``because you expect there to be a check around where those strings are.''

\paragraph{Review program metadata (RQ2)} Some \REs{} looked at information beyond the binary or execution trace, such as the file metadata (N=3), any additional resources loaded (N=3) (e.g., images or additional binaries), function size (N=2), history of recent changes (N=1), where vulnerabilities were found previously (N=1), and security mitigations used (N=1) (e.g., DEP or ASLR). This information gives further insights into program functionality and can help \REs{} know what not to look for. \pfour{} said ``I've been burned in the past. You kind of end up down a long rabbit hole that you have to step completely back from if you don't realize these things\ldots{}For example, for PIE [Position Independent Executables] there has to be some sort of program relative read or write or some sort of address disclosure that allows me to defeat the randomization. So that's one thing to look for early on.'' 

\paragraph{Malware analysts perform \overview{} after unpacking (RQ2)} Many malware binaries are stored in obfuscated form and only deobfuscated at execution time to complicate \RE{}. This is commonly referred to as \emph{packing}. Therefore, \REs{} must first unpack the binary before strings and imported APIs become intelligible (N=2). However, once unpacking is performed and the binary is in a readable state, \REs{} perform the same \overview{} analyses described above (N=2). 

\paragraph{\Overview{} is unique to \RE{} (RQ3)} In most other program comprehension tasks, the area of code to focus on is known at the outset based on the error being debugged~\cite{zeller2009programs} or the functionality being modified or updated~\cite{DevsSeekKo2006,Robillard2004}. Additionally, developers performing program comprehension tasks typically have access to additional resources, such as documentation and the original developers, to provide high-level understanding~\cite{Roehm2012}, making \overview{} analyses unnecessary.

\subsection{\SKIMMING{} (RQ1)}
Based on findings from their \overview{}, \REs{} next shift their attention to program sub-components, searching for insights into the ``how'' of program functionality. By focusing on sub-components, \skimming{} allows \REs{} to quickly identify or rule out hypotheses and refine their view of the program. \peight{} explained that he scanned the code instead of reading line-by-line, saying, ``I'm going through it at a high level, because it's really easy to get caught in the weeds when there could be something much better to look at.'' The middle column of Figure~\ref{fig:tiers} gives an overview of this analysis phase.


\begin{figure}[t]
    \includegraphics[width=\columnwidth]{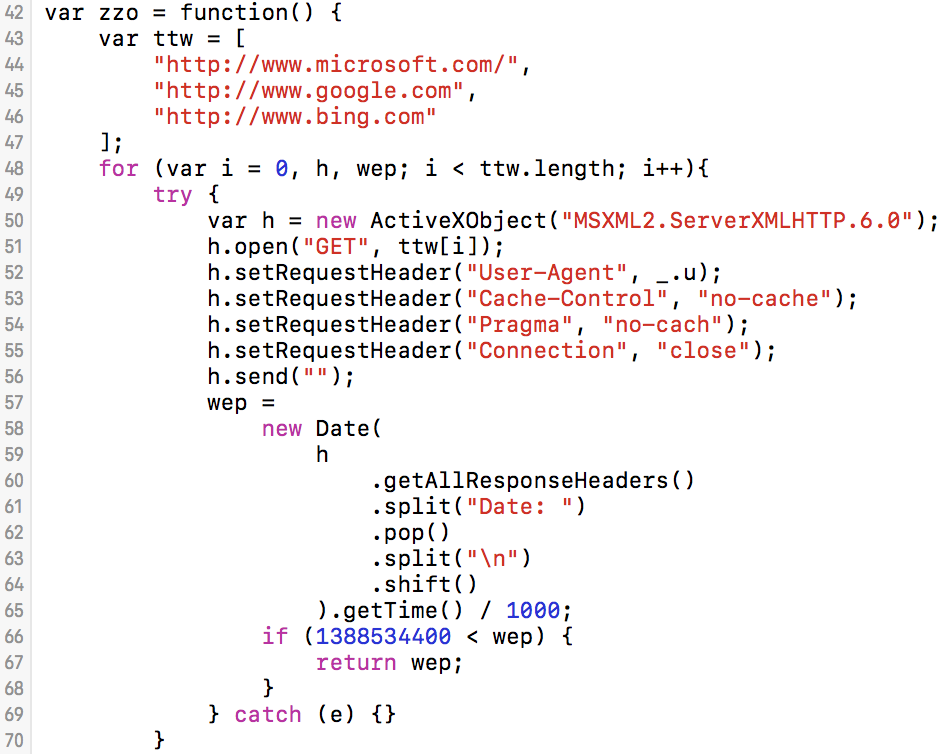}
    \caption{Screenshot of botnet code investigated by \peleven{}, which performs a network connectivity check. This provides an example of API calls and strings recognized during \skimming{} giving program functionality insights.}
    \label{fig:beaconsequence}
        \vspace{-.4cm}
\end{figure}

\paragraph{Scan for many \beacons{} (RQ2)} Most commonly, \REs{} scan through functions or code segments prioritized in the \overview{} (N=15), looking for a variety of \beacons{} indicating possible behaviors. These include APIs (N=15), strings (N=15), constants (N=11), and variable names (N=11). For example, while investigating a piece of malware, \ptwo{} saw \lstinline{GetProcAddress} was called. This piqued his interest because ``it's a very common function for obfuscation\ldots{}it's likely setting up an alternate input table'' to hide obviously malicious calls from an \RE{} looking only at the standard import table.

\REs{} infer program behaviors both from individual instances (N=16) and specific sequences (N=12) of these items. For example, while reverse engineering the code in Figure~\ref{fig:beaconsequence}, \peleven{} first scanned the strings on lines 44-46 and recognized them as well-known websites, generally reachable by any device connected to the Internet. He then looked at the API calls and strings on lines 51-56 and said that ``it's just trying to make a connection to each of those [websites].'' By looking at the constant checked on line 66, he inferred that ``if it's able to make a connection, it's going to return a non-zero value [at line 66].'' Putting this all together and comparing to past experience, \peleven{} explained, ``usually you see this activity if something is trying to see if it has connectivity.''

\REs{} also make inferences from less obvious information. Many review control-flow structures (N=13) for common patterns. When studying a router's firmware, \peight{} noticed an assembly code structure corresponding to a switch statement comparing a variable to several constants. From this, he assumed that it was a  ``comparison between the device's product ID and a number of different product IDs. And then it's returning different numbers based off that. So it looks like it's trying to ascertain what product it is and then doing something with it,'' because he has ``seen similar behavior before where firmware is written in generically.'' Other \REs{} consider the assembly instructions chosen by the compiler (N=8) or function prototypes (N=5) to determine the data types of variables. \ptwo{} explained, ``It is very important to understand\ldots{}how compilers map code to the actual binary output.'' As an example, he pointed out instructions at the start of a function and said, ``that's just part of saving the values\ldots{}I can safely skip those.'' Then he identified a series of registers and observed ``those are the function's arguments\ldots{}after checking the codebase of FreeBSD, I know the second argument is actually a packed structure of arguments passed from outside the kernel. This is [the data] we control in this function context.'' Finally, \REs{} consider the code's relation to the overall program flow (N=6). For example, \peight{} identified a function as performing ``tear down'' procedures---cleaning up the state of the program before terminating---because it ``happened after the main function.''



\begin{figure}
    \includegraphics[width=\columnwidth]{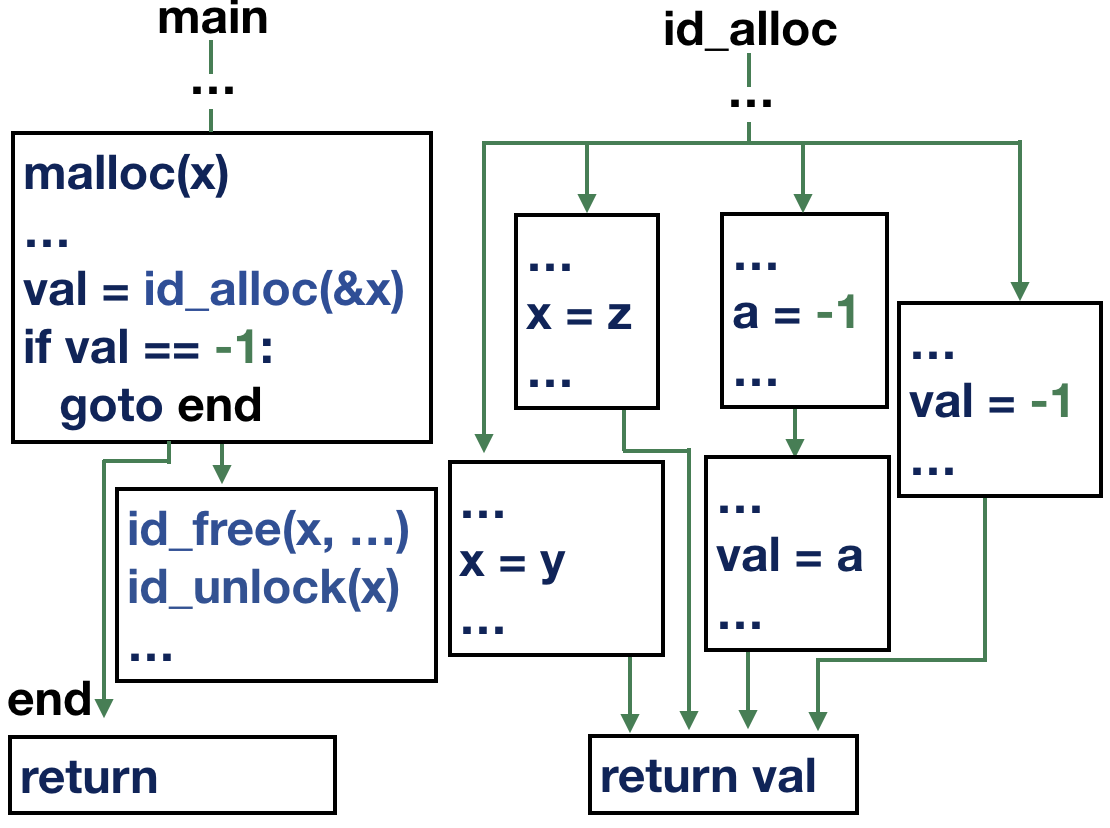}
    \caption{Program investigated by \ptwo{} to determine whether he could trigger an undefined memory read. The code has been converted to a pseudo-code representation including only relevant lines. It shows the control flow graph for two functions: \lstinline{main} and \lstinline{id_alloc}. Rectangles represent basic blocks, and arrows indicate possible control flow paths.}
    \label{fig:controlflowdataflow}
        \vspace{-.4cm}
\end{figure}

\paragraph{Focused on specific data-flow and control-flow paths (RQ2)} Some \REs{} also scanned specific data- (N=8) and control-flow (N=7) paths, only considering instructions affecting these paths. These analyses were commonly used to understand how a function's input (N=7) or output (N=4) is used and whether a particular path is realizable (N=4). For example, while reviewing the program summarized in Figure~\ref{fig:controlflowdataflow}, \ptwo{} asked whether a control-flow path exists through \lstinline{id_alloc} in which \lstinline{x} is not written. Memory for \lstinline{x} is allocated before the \lstinline{id_alloc} call and read after, so if such a path is possible, ``we can have it read from undefined memory.'' To answer this question, \ptwo{} scanned each control flow path through the function from the bottom of the graph up. If he saw a write to \lstinline{x}, he moved on to the next path. This check invalidated the first two control-flow paths (counting left-to-right) in Figure~\ref{fig:controlflowdataflow}. Additionally, in \lstinline{main}, the program exits if the return value of \lstinline{id_alloc} is \lstinline{-1}. Thus his next step was to check the data flow to \lstinline{id_alloc's} return value to see whether it was set to \lstinline{-1}. He found the return value was set to \lstinline{-1} in both remaining control-flow paths, indicating it was not possible to read from undefined memory.

\paragraph{The diversity of \beacons{} represents a second difference from program comprehension (RQ3)} While program comprehension research has identified several similar \beacons{} (API calls, strings, variable names, sequences of operations, and constants~\cite{SchemaBasedDetienne1990,CognitivePCArunachalam1996,DevsSeekKo2006,StimulusStructuresPennington1987}), developers have been shown to struggle when variable names and other semantic information are obfuscated~\cite{SchemaBasedDetienne1990}. However, \REs{} adapt to the resource-starved environment and draw on additional \beacons{} (i.e., control flow structures, compiler artifacts, and program flow). 

\subsection{\INDEPTH{} (RQ1)}
\label{sec:indepth}
Finally, when \REs{} identify a specific question or hypothesis, they shift to \indepth{}: setting up small experiments, varying program inputs and environmental conditions, and considering the program's behavior in these states to find a concrete answer or prove whether specific hypotheses hold. This phase's results are fed back into \skimming{}, to refine high-level hypotheses and the \RE{}'s interpretation of observed beacons. Again, \REs{} rely on a wide range of methods for this analysis. 


\paragraph{Execute the program (RQ2)} In most cases, \REs{} validate their hypotheses by running the code under specific conditions to observe whether the expected behavior occurs (N=13). They may try to determine what value a certain variable holds at a particular point (e.g., input to a function of interest) under varying conditions (N=13) or whether user input flows to an unsafe function (N=9). For example, after reviewing the data-flow path of the program's arguments, \pthree{} hypothesized that the program required two input files with a specific string in the first line to allow execution to reach potentially vulnerable code. To test this hypothesis, she ran the program in a debugger with the expected input and traced execution to see the state of memory at the potentially vulnerable point.

While running the program, \REs{} gather information in a variety of ways. Most execute the code in a debugger (N=12) to probe memory and have full control over execution. Some use other tools like packet capturers and file monitors to observe specific behaviors (N=8). In some cases, \REs{} manipulate the execution environment by dynamically changing registry values (N=7) or patching the binary (N=5) to guide the program down a specific path. As an example, while analyzing malware that ``checks for whether it is being run in a debugger,'' \psixteen{} simply changes the program ``so that the check will always just return false [not run in debugger].''

Finally, some \REs{} fuzz program inputs to identify mutation-specific behavior changes. In most cases, fuzzing is performed manually (N=6), where the \RE{} hand-selects mutations. Automation is used in later stages, once a good understanding of the program is established (N=1). \peight{} explained, ``I wait until I have a good feel for the inputs and know where to look, then I patch the program so that I can quickly pump fuzzed inputs from angr~\cite{angr} into the parts I care about.''

\paragraph{Compare to another implementation (RQ2)} Some \REs{} chose to re-write code segments in a high-level language based on the expected behavior (N=8) or searched for public implementations (e.g., libraries) of algorithms they believed programs used (N=5). They then compared the known implementation's outputs with the subject program's outputs to see if they matched. For example, once \pten{} recognized the encryption algorithm he was looking at was likely Blowfish, he downloaded an open-source Blowfish implementation. He first compared the open-source code's structure to the encryption function he was reviewing. He then ran the reference implementation and malware binary on a file of all zeros saying, ``we can then verify on this sample data whether it's real Blowfish or if it's been modified.''

\paragraph{Read line-by-line only for simple code or when execution is difficult (RQ2)} Finally, \REs{} resorted to reading the code line-by-line and mentally tracking the program state when other options became too costly (N=9). In some cases, this occurred when they were trying to answer a question that only required reading a few, simple lines of code. For example, \pfive{} described a situation where he read line-by-line because he wanted to fully understand a small number of specific checks, saying, ``After Google Project Zero identified some vulnerabilities in the system, the developers tried to lock down that interface by adding these checks. Basically I wanted to figure out a way to bypass these specific checks. At this point I ended up reading line-by-line and really trying to understand the exact nature of the checks.'' While no participants quantified the number of lines or code complexity they were willing to read line-by-line, we did not observe any participants reading more than 50 lines of code. Further, this determination appeared goal- and participant-dependent, with wide variation between participants and even within individual participants' own processes, depending on the current experiment they were carrying out.

\REs{} also chose to read line-by-line instead of running the program when running the program would require significant setup (e.g., when using an emulator to investigate uncommon firmware like home routers). \pnine{} explained, ``The reason I was so IDA [disassembler] heavy this time is because I can't run this binary. It's on a cheap camera and it's using a shared memory map. I mean, I could probably run this binary, but it's going to take a while to get [emulation] set up.''

During this line-by-line execution, a few \REs{} said they used symbolic execution to track inputs to a control flow conditional of interest (N=2). \pthree{} explained, ``I write out the conditions to see what possible states there are. I have all these variables with all these constraints through multiple functions, and I want to say for function X, which is maybe 10 deep in the program, what are the possible ranges for each of these variables?'' In both cases, the \REs{} said they generally performed this process manually, but used a tool, such as Z3, when the conditions became too complicated. As \pthree{} put it, ``It's easier if you can just do it in your brain of course, but sometimes you can't\ldots{} if there are 10 possibilities or 100 possibilities, I'll stick it in a SAT solver if I really care about trying to get past a barrier [conditional].''

\paragraph{\Beacons{} are still noticed and can provide shortcuts (RQ2)} While \REs{} focus on answering specific questions in this phase, some also notice \beacons{} missed in prior analyses. If inferences based on these \beacons{} invalidated prior beliefs, \REs{} quickly stop \indepth{} that becomes moot. For example, while \pfour{} was reverse engineering a card-game challenge binary, he decided to investigate a \lstinline{reset} function operating on an array he believed might be important. There were no obvious \beacons{} on initial inspection and there were only a few instructions, so he decided to read line-by-line. However, he quickly recognized two constants that allowed him to infer functionality. He saw that ``it's incrementing values from 0 to 51. So at this point, I'm thinking it's a deck of cards. And then it has this variable hold. Hold is a term for poker, and it sets 0 to 4.'' Once he realized what these variables were, he decided he had sufficient information to stop analyzing the function, and he moved back to the calling function to resume \skimming{}.

\paragraph{Simulation methods mostly overlap with program comprehension (RQ3)} Most of the methods described above, including using a debugger and reading code line-by-line, are found in the program comprehension literature. However, comparing program execution to another implementation appears unique to \REs{}. As in \skimming{}, this extra method is likely necessitated by the additional complexity inherent in an adversarial environment.

\section{Results: Cross-phase Trends}
\label{sec:trends}
\begin{figure}
    \includegraphics[width=\columnwidth]{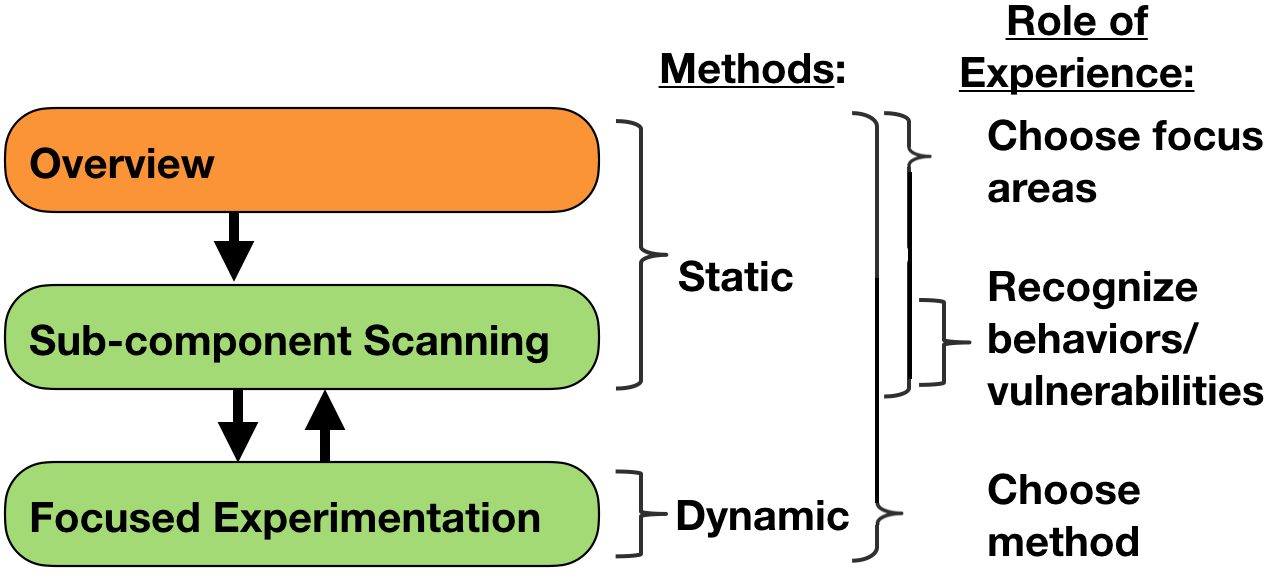}
    \caption{Overview of the analysis phases and trends observed across them. The arrows shown between the phases indicates information flow. The brackets indicate which phases the adjacent item is relevant to.}
    \vspace{-.3cm}
    \label{fig:trends}
\end{figure}

In addition to the phases themselves, we observed several cross-phase trends in our participants' \RE{} approaches, which we discuss in this section. This includes both answers to our research questions which were not unique to a specific phase and additional observations regarding tool usage which inform future tool development. Figure~\ref{fig:trends} includes some of these trends as they interact with the phases.


\paragraph{Begin with static methods and finish with dynamic (RQ2)} Most of the simulation methods described in the first two analysis phases focused on static program representations, i.e., the binary or decompiled code. In contrast, \indepth{} was mainly performed dynamically, i.e., by running the program. Reverse engineers typically make this switch, as \pfive{} stated, ``because this thing is so complex, it's hard to trace the program flow [statically], but you can certainly tell when you analyze an [execution] trace. You could say this was hit or this wasn't hit.'' However, \REs{} sometimes choose not to switch when they perceive the switch to be difficult. \pfifteen{} explained ``[switching] was a little daunting to me. I just wanted to work in this environment I'd already set up.'' 


Unfortunately, in most cases, switching contexts can be difficult because \REs{} have to manually transfer information back and forth between static and dynamic tools (e.g., instructions or memory states) (N=14). To overcome this challenge, some \REs{} opened both tools side-by-side to make comparisons easier (N=4). 
For example, \peight{} opened a debugger in a window next to a disassembler and proceeded to step through the \lstinline{main} function in the debugger while following along in the assembly code. As he walked through the program, he regularly switched between the two. For example, he would scan the possible control-flow paths in the disassembler to decide which branch to force execution down and the necessary conditions would be set through the debugger. Whenever he came across a specific question that could not be answered just by scanning, he would switch to the debugger. Because he stepped through the program as he scanned, he could quickly list register values and relevant memory addresses to get concrete variable values.

\paragraph{Experience and strategy guide where to look in the first two phases (RQ1)} Initially, \REs{} have to make decisions about which metadata to look at, e.g., all strings and APIs or specific subsets, (N=4) and what inputs to provide to exercise basic behaviors (N=2). Once they run their \overview{} analyses, they must determine which outputs (strings, APIs, or UI elements) are relevant to their investigation (N=16) and in what order to process them (N=11). Reverse engineers first rely on prior experience to guide their actions (N=14). \pfour{} explained that when he looks for iPhone app vulnerabilities, he has ``a prioritized list of areas [APIs] I look at...it's not a huge list of things that can go horribly wrong from a security standpoint when you make an iPhone app...So, I just go through my list of APIs and make sure they're using them properly.'' If \REs{} are unable to relate their current context to prior experience, then they fall back on basic strategies (N=16) such as looking at the largest functions first. \pthree{} said, ``If I have no clue what to start looking at...I literally go to the function list and say the larger function is probably interesting...as long as I can distinguish the actual code versus library code, this technique is actually pretty useful.'' Similarly, \REs{} employ heuristics to decide which functions not to investigate. For example, \psixteen{} said, ``If the function is cross-referenced 100 times, then I will avoid it. It's probably something like an error check the compiler added in.''

In \skimming{}, experience plays an even more important role. As in the previous analysis phase, \REs{} must decide which data- (N=8) and control-flow paths (N=7) to consider. Again, this is done first by prior experience (N=6) and then by simple strategies (N=4). As they perform their analyses, \REs{} must also determine potential hypotheses regarding program functionality (N=16) and possible vulnerabilities (N=9)---exploitable flaws in the case of vulnerability discovery, or signaturable behaviors for malware analysis. In most cases, these determinations are made by recognizing similarities with previous experiences (N=15). For example, when \peight{} saw a function named \lstinline{httpd_ipc_init}, he recognized this might introduce a vulnerability, saying, ``IPC generally stands for inter-process communication, and many router firmwares like this set up multiple processes that communicate with each other. If it's doing IPC through message passing, then that opens up the attack surface to anything that can send messages to this httpd binary.'' If the \RE{} is unable to generate hypotheses based on prior experience, they instead make determinations based on observed behaviors (N=16), obtained via more labor intensive investigation of the program execution or in-depth code review.



\paragraph{Experience used to select analysis method throughout (RQ1)} There were typically multiple ways to answer a question. The most common example, as discussed in Section~\ref{sec:indepth}, was deciding between executing the program or reading line-by-line during \indepth{} (N=9). Similar decisions occurred in the other phases. For example, some \REs{} choose to simply skip the \overview{} phase all together and start with the \lstinline{main} function (N=5) whenever, as \pthree{} said, ``it's clear where the actual behavior starts that matters.'' 

\REs{} also decide the granularity of analysis, weighing an approximation's benefits against the inaccuracy introduced (N=5). For example, several participants discussed choosing to use a decompiler to make the code easier to read, knowing that the decompilation process introduces inaccuracies in certain circumstances. \pfour{} said, ``I actually spend most of my time in Hex-Rays [decompiler]. A few of my friends generally argue that this is a mistake because Hex-Rays can be wrong, and disassembly can't be. And this is generally true, but Hex-Rays is only wrong in specific ways.'' Further, because these are explicit decisions, \REs{} are also able to recognize situations where the inaccuracies are common and can switch analysis granularities to verify results (N=5). For example, when using a decompiler, the \RE{} has some intuition regarding what code should look like. \pfour{} explained, ``I've had many situations where I think this looks like an infinite loop, but it can't be. It's because Hex-Rays is buggy. Basically, in programming, no one does anything all that odd.''

\paragraph{Preferred tools presented output in relation to the code (RQ2)} In almost all cases, the tools \REs{} choose to use provide a simple method to connect results back to specific lines of code (N=16). They choose to list strings and API calls in a disassembler (N=15), such as IDA, which shows references in the code with a few clicks, as opposed to using the command-line \lstinline{strings} command (N=0). Similarly, those participants who discussed using advanced automated analyses, i.e., fuzzing (N=1) and symbolic execution (N=1), reported using them through disassembler plugins which overlaid analysis results on the code (e.g., code coverage highlighting for fuzzing). \pthree{} used Z3 for symbolic execution independently of the code, supplying it with a list of possible states and manually interpreting its output with respect to the program. However, she explained this decision was made because she did not know a tool that presented results in the context of the code that could be used with the binary she was reversing. She said, ``The best tool for this is PAGAI\ldots{}If you have source it can give you ranges of variables at certain parts in a program, like on function loops and stuff.'' Specifically, PAGAI lets \REs{} annotate source code to define variables of interest and then presents results in context of these annotations~\cite{Henry2012Pagai}. 

\paragraph{Focused on improving readability (RQ2)} Throughout, \REs{} pay special attention to improving code readability by modifying it to include semantic information discovered during their investigation. In most cases, the main purpose of tools \REs{} used was to improve code readability (N=9). Many \REs{} used decompilers to convert the assembly code to a more readable high-level language (N=9), or tools like IDA's lumina server~\cite{lumina} to label well-known functions (N=2). Additionally, most \REs{} performed several manual steps specifically to improve readability, such as renaming variables (N=14), taking notes (N=14), and reconstructing data structures (N=8). \pone{} explained the benefit of this approach when looking at a file reading function by saying, ``It just says call DWORD 40F880, and I have no idea what that means\ldots so, I'll just rename this to read file\ldots [now I know] it's calling read file and not some random function that I have no idea what it is.'' Taking notes was also useful when several manipulations were performed on a variable. For example, to understand a series of complex variable manipulations, \pfive{} said ``I would type this out. A lot of times I could just imagine this in my head. I think usually I can hold in my head two operations...If it's anything greater than that I'll probably write it down.''


\paragraph{Online resources queried to understand complex underlying systems (RQ2)} Regarding external resources, \REs{} most often reference system and API documentation (N=10). They reference this documentation to determine specific details about assembly opcodes or API arguments and functionality. They also reference online articles (N=4) that provide in-depth breakdowns of complicated, but poorly documented system functions (e.g., memory management, networking, etc.). When those options fail, some \REs{} also reference question-answering sites like StackOverflow (N=4) because ``sometimes with esoteric opcodes or functions, you have to hope that someone's asked the question on StackOverflow because there's not really any good documentation'' (P3). Many participants also google specific constants or strings they assume are unique to an algorithm (N=7). P10 explained, ``For example, MD5 contains an initialization vector with a constant. You just google the constant and that tells you the algorithm.''

\section{Discussion}
\label{sec:discussion}

Our key finding is the identification and description of a three-phase \RE{} process model, along with cross-phase trends in \REs{}' behaviors. This both confirms and expands on prior work, which described an \RE{} model of increasingly refined hypotheses~\cite{Bryant2012}. We demonstrate a process of hypothesis generation and refinement through each phase, but also show the types of questions asked, hypotheses generated, actions taken, and decisions made at each step as the \RE{} expands their program knowledge.

Our model highlights components of \RE{} for tool designers to focus on and provides a language for description and comparison of \RE{} tools. Building on this analysis model, we propose five guidelines for \RE{} tool design. For each guideline, we discuss the tools closest to meeting the guideline (if any), how well it meets the guideline, and challenges in adopting the guideline in future tool development. Table~\ref{tab:guidelines} provides a summary, example application, and challenges for each guideline. While these guidelines are drawn directly from our findings, further work is needed to validate their effectiveness.

\begin{table*}
\footnotesize
\centering
\begin{threeparttable}
\footnotesize
\begin{tabular}{l | p{77mm} | p{77mm}}
 & \multicolumn{1}{c |}{\textbf{Reverse Engineering Tool Design Guidelines}} & \multicolumn{1}{c}{\textbf{Example Application}}\\
  \toprule
 G1 & \textbf{\Gone{}} \newline Reverse engineering tools should be designed to facilitate each analysis phase: \overview{}, \skimming{}, and \indepth{}. & \textbf{IDAPro~\cite{ida}, BinaryNinja~\cite{binaryninja}, Radare2~\cite{radare}} \newline Provide platforms for \REs{} to combine analyses, but previously lacked thorough \RE{} process model to guide analysis development and integration. \\
 \hline
 G2 & \textbf{\Gtwo{}} \newline Integrate analysis interaction into the disassembler or decompiled code view to support tool adoption & \textbf{Lighthouse~\cite{lighthouse}} \newline Highlights output in the context of code, but does not support input in code context. \\
 \hline
 G3 & \textbf{\Gthree{}} \newline Static and dynamic analyses should be tightly coupled so that users can switch between them during exploration. & \textbf{None we are aware of} \newline We do not know of any complex analysis examples. This is possibly due to challenges with visualization and incremental analysis. \\
  \hline
 G4 & \textbf{\Gfour{}} \newline When multiple options for analysis methods or levels of approximation are available, ask the user to decide which to use. & \textbf{Hex-rays decompiler~\cite{hexray}} \newline Minimally applies G4 by giving users a binary option of a potentially imprecise decompiled view or a raw disassembly view. \\
  \hline
  G5 & \textbf{\Gfive} \newline Infer semantic information from the code where possible and allow users to change variable names, add notes, and correct decompilation to improve readability. & \textbf{DREAM++ decompiler~\cite{Yakan2016DREAM++}} \newline Provides significantly improved decompiled code readability through several heuristics, but is limited to a preconfigured set of readability transformations. \\
  \bottomrule
\end{tabular}
\end{threeparttable}
\caption {Summary of guidelines for \RE{} tool interaction design.}
\label{tab:guidelines}
\vspace{-5mm}
\end{table*}

\paragraph{G1. \Gone{}} The most obvious conclusion is that \RE{} tools should be designed to mesh with the three analysis phases identified in Section~\ref{sec:tiers}. This means \REs{} should first be provided with a program overview for familiarization and to provide feedback on where to focus effort (\overview{}). As they explore sub-components, specific slices of the program (\beacons{} and data/control-flow paths) should be highlighted (\skimming{}). Finally, concrete, detailed analysis information should be produced on demand, allowing \REs{} to refine their program understanding (\indepth{}). 

While this guideline is straightforward, it is also significant, as it establishes an overarching view of the \RE{} process for tool developers. Because current \RE{} tool development is ad-hoc, tools generally perform a single part of the process and leave the \RE{} to stitch together the results of several tools. 
G1 provides valuable insights to single-purpose tool developers by identifying how they should expect their tools to be used and the input and output formats they should support. Additionally, with the growing effort to produce human-assisted vulnerability discovery systems~\cite{chess}, G1 shows when and how human experts should be queried to support automation.

The closest current tools to fulfilling G1 are popular reverse engineering platforms such as IDAPro~\cite{ida}, BinaryNinja~\cite{binaryninja}, and Radare~\cite{radare}, which provide disassembly and debugger functionality and support user-developed analysis scripts. These tools allow \REs{} to combine different analyses (N=16). However, due to these tools' open-ended nature and the lack of a prior \RE{} process model, there are no clear guidelines for script developers, and users often have to perform significant work to find the right tool for their needs and incorporate it into their process.


\paragraph{G2. \Gtwo{}}
We found that most \REs{} only used tools whose interactions were tightly coupled with the code. This suggests that tool developers should place a high priority on allowing users to interact directly with (disassembled or decompiled) code. The best example of this we observed was given by \pfive{} in the code-coverage visualization plugin Lighthouse, which takes execution traces and highlights covered basic blocks in a disassembler view~\cite{lighthouse}. It also provides a ``Boolean query where you can say only show me covered blocks that were covered by this trace and not that trace, or only show blocks covered in a function whose name matches a regular expression.'' However, Lighthouse does not fully follow our recommendation, as there is no way to provide input in the context of the code. For example, the user might want to determine all the inputs reaching an instruction to compare their contents.  However, this is not currently possible in the tool.

\paragraph{G3. \Gthree{}}
We found that almost all participants switched between static and dynamic program representations at least once (N=14). This demonstrates tools' need to consider both static and dynamic information, associate relevant components between static and dynamic contexts, and allow \REs{} to seamlessly switch between contexts. For example, \pfour{} suggested a dynamic taint analysis tool that allows the user to select sinks in the disassembler view, run the program and track tainted instructions, then highlight tainted instructions again in the disassembler view. This tool follows our suggested guideline, as it provides results from a specific execution trace, but also allows the user to contextualize the results in a static setting. 

We did observe one participant using a tool which displayed the current instruction in the disassembly view when stepping through the code in a debugger, and there have been several analyses developed which incorporate static and dynamic data~\cite{stephens2016driller,hallerDowser2013,WangTaintScope2010,Drewry2007,wong2016intellidroid,ZhengSmartDroid2012}. However, we are unaware of any more complex analyses that support user interaction with both static and dynamic states. Following G3 requires overcoming two difficult challenges.  First, the analysis author must determine how to best represent dynamic information in a static setting and vice versa. This requires careful design of the visualization to ensure the user is provided relevant information in an interpretable manner. Second, we
speculate that incremental program analyses (such as those of Szabo et al.~\cite{Szabo:2016}) may be necessary in
this setting to achieve acceptable performance compared to current batch-oriented tools.

%
%

\paragraph{G4. \Gfour{}}
Throughout the \RE{} process, \REs{} choose which methods to use based on prior experiences and specific needs, weighing the method's benefit against any accuracy loss (N=5). These tradeoff decisions are inherent in most analyses. Therefore, we recommend tool designers leverage \REs{}' ability to consider costs and also recognize instances where the analysis fails. This can be done by allowing \REs{} to select the specific methods used and tune analyses to fit their needs. One example we observed was the HexRays decompiler~\cite{hexray}, which allows users to toggle between a potentially imprecise, but easier to read, decompiled program view and the more complex disassembled view. This binary choice, though, is the minimum implementation of G4, especially when considering more complex analyses where the analysis developer must make several nuanced choices involving analyses such as context, heap, and field sensitivity~\cite{Smaragdakis:2011}. This challenge becomes even more difficult if the user is allowed to mix analysis precision throughout the program, as static analysis tools generally use uniform analysis sensitivity.
However, recent progress indicates that such hybrid analyses are beginning to receive attention~\cite{Kastrinis:2013,Gilray:2016}.


\paragraph{G5. \Gfive{}}
We found most \REs{} valued program readability improvements. Therefore, \RE{} tool designers should allow the user to add notes or change naming to encode semantic information into any outputs. Further, because annotation is such a common behavior (N=14), tools should learn from these annotations and propagate them to other similar outputs. The best example of a tool seeking to follow this recommendation is the DREAM++ compiler by Yakdan et al.~\cite{Yakan2016DREAM++}. DREAM++ uses a set of heuristics derived from feedback from \REs{} to provide semantically meaningful names to decompiled variables, resulting in significant readability improvements. One improvement to this approach might be to expand beyond DREAM++'s preconfigured set of readability transformations by observing and learning from developer input through renaming and annotations. This semantic learning problem poses a significant challenge for the implementation of G5, as it likely requires the analysis to consider minor nuances of the program context.


\paragraph{\RE{} tool designers should consider the exploratory visual analysis (EVA) literature} In addition to the guidelines drawn directly from our results, we believe \RE{} tool designers can draw inspiration from EVA. EVA considers situations where analysts search large datasets visually to summarize their main characteristics. Based on a review of the EVA literature, Battle and Heer define a process similar to the one we observed \REs{} to perform, beginning with a high-level overview, generating hypotheses, and then iteratively refining these hypotheses through a mix of scanning and detailed analysis~\cite{2019-exploratory-visual-analysis}. Further, Shneiderman divided EVA into three phases, similar to those we suggest, with his Visual Information Seeking Mantra: ``Overview first, zoom and filter, then details-on-demand''~\cite{Shneiderman1996}. While techniques from this field likely cannot be applied as-is due to differences in the underlying data's nature, these similarities suggest insights from EVA could be leveraged to guide similar development in \RE{} tools, including methods for data exploration~\cite{Heer2012,Perer2008,Kalinin2014,Siddiqui2016}, interaction~\cite{Yi2007,Heer2008,JankunKelly2007,Pike2009}, and predicting future analysis questions~\cite{Battle2016,Gotz2009,Dimitriadou2014,Vartak2015}. 

\section{Conclusion}

Our goal is to carefully model \REs{}' processes, in order to 
support better design of \RE{} tools. To do this, we conducted  
a semi-structured observational interview study of 16 professional 
\REs{}. We found that \RE{} involves three distinct phases: \overview{},
\skimming{}, and \indepth{}. Reverse engineers work through a 
program using a variety of manual and automated approaches 
in each of these phases, often using a combination of methods 
to accomplish a specific task (e.g., a static analysis alongside a 
debugger). In the first two phases (\overview{} and \skimming{}), 
\REs{} typically use static techniques (e.g., looking at a control-flow
graph), but switch to using dynamic techniques (e.g., debugging or
dynamic analysis) in the last phase (\indepth{}). Based on our
results, we proposed five design guidelines for \RE{}
tools. We believe our model will help in the design and 
development of \RE{} tools that more closely match the
\RE{} process.




\section*{Acknowledgments}

We thank Kelsey Fulton and the anonymous reviewers for their helpful feedback; BinaryNinja, the two bug-bounty 
platform companies, and the many CTF teams that supported our recruitment efforts; and Jordan 
Wiens for providing valuable insights into the world of reverse engineering. This research was supported in part 
by a UMIACS contract under the partnership between the University of Maryland and DoD, and by a Google 
Research Award.



%
\bibliographystyle{IEEEtran}
{\footnotesize
\bibliography{IEEEabrv,paper}}

\appendix

\section{Interview protocol}
\label{appendix:protocol}
\subsection{App Background}
To begin our discussion, I want you to think of a program that you recently reverse engineered.
\begin{enumerate}
 
\item What was the name of the program?  [If they're not comfortable telling the name, there are a few additional cues below]
\begin{enumerate}
\item What type of functionality did the app provide? [Exs: Banking, Messaging, Social Media, Productivity]
\item Approximately, how many lines of code or number of classes did the app have?
\end{enumerate}
\item Why were you investigating this program?
\item Approximately, how long did you spend reverse engineering this app?
\item What tools did you use for your reverse engineering process?  [Exs: IDAPro, debugger, fuzzer]
\item Did you reverse engineer this app with other people?
\begin{enumerate}
\item (If yes) how did you divide up the work?
\end{enumerate}
\end{enumerate}

\subsection{Reverse Engineering Process}
Next, we'll talk about this app in more detail.  If possible, I would like you to open the program you searched the same way you did when you first started investigating it.  If you would like to share your screen with me, that would be helpful for providing context, however, this is not necessary.  Primarily, I want you to open everything on your computer to help you remember the exact steps you took when you searched the program. 

[If they do share their screen]
Also, if you are comfortable, I would like to record this screen sharing session, so that we have a later reference.

Please walk me through how you searched the program.  As you go through your process, please explain every step you took, even if it was not helpful toward your eventual goal.  For example, if you decided to reverse engineer a specific class, but realized it was not relevant to your search after reading the code, we would still like to know that you performed this step.
[a few cueing questions are provided below to guide the conversation]
\begin{enumerate}
\item Where did you start?
\item What questions did you ask? How did you answer these questions?
\end{enumerate}

\subsection{Items of Interest}

\paragraph{Decision Points}
[Every time the participant had to decide between one or more actions during their process. Ex: Where to start? What test cases to try? Which path to go down first? When to inspect a function?]
\begin{enumerate}
\item Record the decision that was made
\item How did you make this decisions?  Explain your thought process
\end{enumerate}

\paragraph{Hypotheses}
[Every time the participant states a question they have to answer or makes a conjecture about what they think the program (or component) does. Ex: X class performs Y function. X data is transmitted off device, it's using Y encryption]
\begin{enumerate}
\item Record the hypothesis or question asked
\item Why did you think this could be the case?
\item How did they (in)validate this hypothesis?
\end{enumerate}

\paragraph{Beacons}
[Every time the participant states recognizing the functionality of some code without actually stepping through it. That is, they are able to notice some pattern in the code and make some deductions about functionality based on this]
\begin{enumerate}
\item Record the beacon that was noticed
\item Why did this stand out to you? How were you able to recognize it?
\item How did you know that it was X instead of something else?
\end{enumerate}

\paragraph{Simulation}
[Every time the participant discusses looking at the code to determine how it works]
\begin{enumerate}
\item Record how they investigate the code.
\begin{enumerate}
\item (If Automation) Do you use a custom tool or something open source/purchased?
\begin{enumerate}
\item (If not custom) What tool do you use?
\begin{enumerate}
\item Does this tool provide the results you would want or does it fall short in some way? [Ex: I actually want output X, but I get Y, so I need to do these steps to get to X]
\end{enumerate}
\end{enumerate}
\item Is this generally the approach you use? 
\begin{enumerate}
\item (If no) Why here and not in other cases?
\item (If yes) What advantage do you think this approach has over other manual/automated investigation?
\end{enumerate}
\end{enumerate}
\item Please describe what's going on in your head or the automation?
\begin{enumerate}
\item What are the inputs and outputs?
\item When do you know when to stop?
\end{enumerate}
\end{enumerate}

\paragraph{Resources}
[Every time the participant discusses referencing some documentation or information source external to the code]
\begin{enumerate}
\item Record what resource they used
\item Do you regularly consult this resource for information?
\item What do you think the benefit of this resource is over other sources of information? [Exs: Language documentation, Stack Overflow, internal documentation]
\end{enumerate}

\section{Survey questionnaire}
\label{appendix:survey}
\begin{enumerate}
\item Please specify the gender with which you most closely identify.	

\begin{enumerate}
\item Male
\item Female
\item Other
\item Prefer not to answer
\end{enumerate}
\item Please specify your age.
\begin{enumerate}		
\item 18-29
\item 30-39
\item 40-49
\item 50-59
\item 60-69
\item Over 70
\end{enumerate}
\item Please specify your ethnicity. Select all that apply		
\begin{enumerate}
\item White
\item Hispanic or Latino
\item Black or African American
\item American Indian or Alaska Native
\item Asian, Native Hawaiian, or Pacific Islander
\item Other
\end{enumerate}
\item Please specify the highest degree or level of school you have completed
\begin{enumerate}
\item Some high school credit, no diploma or equivalent
\item High school graduate, diploma or the equivalent (for example: GED)
\item Some college credit, no degree
\item Bachelor`s degree
\item Master`s degree
\item Doctoral degree
\end{enumerate}
\item If you are currently a student or have completed a college degree, please specify your field(s) of study (e.g. Biology, Computer Science, etc).

\item Please select the response option that best describes your current employment status.
\begin{enumerate}		
\item Working for payment or profit
\item Unemployed
\item Looking after home/family
\item A student
\item Retired
\item Unable to work due to permanent sickness or disability
\item Other
\item Prefer not to answer
\end{enumerate}

\item Please specify the range which most closely matches your total, pre-tax, personal income specifically from vulnerability discovery in 2017.
\begin{enumerate}		
\item < \$999
\item \$1,000 - \$4,999
\item \$5,000 - \$14,999
\item \$15,000 - \$29,999
\item \$30,000-\$49,999
\item \$50,000-\$74,999
\item \$75,000-\$99,999
\item \$100,000-\$124,999
\item \$125,000-\$149,999
\item \$150,000-\$199,999
\item > \$200,000
\end{enumerate}
\item Prefer not to answer
\item On a scale from 1-5, how would you assess your reverse engineering skill level (1 being a beginner and 5 being an expert)?

\item How many total years of experience do you have with reverse engineering?

\item Please select the range that most closely matches the amount of time you typically spend performing reverse engineering tasks per week.
\begin{enumerate}		
\item <5 hours 
\item 5-10 hours
\item 10-20 hours
\item 20-30 hours
\item 30-40 hours
\item 40+ hours
\end{enumerate}
\item Please select the range that most closely matches the amount of time you typically spend performing non-reverse engineering, technical tasks per week (e.g. software or hardware programming, system administration, network analysis, etc).
\begin{enumerate}		
\item <5 hours 
\item 5-10 hours
\item 10-20 hours
\item 20-30 hours
\item 30-40 hours
\item 40+ hours
\end{enumerate}

\item Please select the range which closely matches the number of software systems you have reverse engineered?
\begin{enumerate}		
\item 0-3
\item 4-6
\item 7-10
\item 11-25
\item 26-50
\item 51-100
\item 101-500
\item 500+
\end{enumerate}
\item Please indicate whether you would be ok with us contacting you regarding future studies even if you are not selected for this study:
\begin{enumerate}
\item I agree to be contacted regarding future studies
\item I do not agree to be contacted regarding future studies
\end{enumerate}
\item Please enter your email address so the we can contact you for the interview, if you are selected.  
\item Your contact information will only be used to invite you to participate in the study. After the study, all records of your contact information will be destroyed unless you indicated above that you agree to be  contacted regarding future stud-ies.
\end{enumerate}

\section{Codebook}
\label{appendix:book}
In this appendix, we list the final codebook used to analyze the content of each interview.  Our codebook was divided into six parts, reflecting our items of interest discussed in Section~\ref{sec:protocol}. For each code, we give a short description where necessary. Some codes were further divided into sub-codes to provide additional specificity to our analysis. We indicate this hierarchical relationship by presenting sub-codes in an indented bulleted list under their parent's code.

\subsection{Hypotheses}
For each hypothesis, we coded both the justification or observation that led to the formulation of a particular hypothesis (\emph{reason}) and type of hypothesis the formed (\emph{type}).

\subsubsection{Reason}
\begin{itemize}
\item Structure - The \RE{} made their inference based on the structure of the data reviewed.  For example, ten digits separated by three dashes is probably a phone number.
\item Observed Behavior - The \RE{} made a determination about program functionality after a full evaluation of the code or execution. That is, they did not rely on outside information to determine the code functionality.
\item Prior Experience - The \RE{} made an inference about program behavior without fully evaluating the code by drawing on similar past experiences.
\end{itemize} 

\subsubsection{Type}
\begin{itemize}
\item Vulnerability - The \RE{} hypothesized that a particular code segment was vulnerability to exploitation.
\item Function - The \RE{} hypothesized what the behavior of a particular code segment was.
\item Data Type or Purpose - The \RE{} hypothesized what the type or purpose of a variable or register was.\end{itemize} 

\subsection{Question}

\begin{itemize}
\item What is the observable behavior of the program? - The \RE{} asked what information could be observed when running the program without using any introspection tools (e.g., debugger, packet capture, etc.).
\item What does the program do for input X? - The \RE{} checks how the program responds when provided with a specific input of interest.
\item How is variable/register/constant X used? - The \RE{} seeks to determine how a specific value of interest is used by the program.  
\item What security controls are being used? - The \RE{} asks what mitigations are in place around the program to prevent exploitation (e.g., ASLR, DEP, etc.)
\item What is the output of function/code X? - The \RE{} seeks to determine the possible output of a function or block of code of interest.
\item What is the possible value of variable/register X at point Y? - The \RE{} seeks to determine all possible values of a specific variable or register at a point of interest in the program.
\item What is the concrete value of variable/register X at point Y? - The \RE{} seeks to determine the value of a variable or register of interest at a specific point in the program given a concrete trace of the program's execution.
\item Where are the strings/names related to X? - The \RE{} seeks to find semantically similar strings and variable or function names in the program related to a concept of interest (e.g., encryption).
\item What input leads to point X being reached? - The \RE{} seeks to determine the specific input that will cause a segment of code of interested to be executed.
\item Can code at point X be reached? - The \RE{} seeks to determine whether it is possible for a segment of code of interest to be executed.
\item How has the program changed over time? - The \RE{} asks what changes to the program's code have been made between the current program version and previous versions.
\item What is the control flow path for input X? - The \RE{} seeks to determine what control flow path through the program is followed when an input of interest is provided.
\item What is the type of variable/register X? - The \RE{} seeks to determine the type of data (e.g., string, integer, pointer, etc.) stored in a variable or register of interest.
\item What is the possible input to function X? - The \RE{} seeks to determine all possible inputs to a function of interest.
\item What is the output of function X used for? - The \RE{} seeks to determine how a functions output is used. For example, is the data transmitted to another device over the Internet and what is the reason for this transmission?
\item What call/uses X (function, string, offset, register)? - The \RE{} asks what other functions call or use a particular item of interest.
\item What does function/code X do? - The \RE{} seeks to determine what the overall behavior of a segment of code or function is.
\item What does function X call? - The \RE{} asks what other functions a function of interest calls.
\end{itemize} 

\subsection{Beacon}

\begin{itemize} 
\item String - In the usual sense, meaning the primitive datatype indicating a series of null terminated characters.
\item API calls - In the usual sense, meaning function calls from external libraries.
\item Low Level Operations - Individual assembly code operations and their parameters (e.g., mov or add).
\item Constants - In the usual sense, meaning primitives in the code that hold a value that does not change.
\item Variable Name - The name of a variable, which can provide hints about the behavior of the program or the purpose of the variable.
\item Operation Sequence -A specific order or sequence of some operations (e.g., API calls, assembly instructions, constants, etc.) that the \RE{} that indicate a particular behavior to the \RE{} on first glance.
\item Comments - Any comments left in the code by the developers. These may be available if the \RE{} has access to source code.
\item Program Metadata - Meta information about the program itself such as the number of lines of code.
\item Function Prototype - In the usual sense, meaning the type that the function returns, its name, and its parameters.
\item UI element - Any element of in the UI of the program.
\item Control Flow - A specific path through the program that is dictated by some decided sequence of conditional branches.
\item Program Flow - The position of a particular function or code segment in relation to the broader order of behaviors. That is, the \RE{} can make inferences about a function or code segment's behavior knowing that it comes before, after, or in concert with other behaviors.
\end{itemize}

\subsection{Simulation Method}
The simulation methods we observed were divided into three groups: dynamic analysis, static analysis, and metadata review. In addition to coding these methods, we also coded interactions of their use.

\subsubsection{Dynamic Analysis}
\begin{itemize}
\item Execute with specific input - Executing the program with input chosen with some specific purpose or idea behind it.
\item Execute in debugger to a certain point - Setting a breakpoint in the debugger and executing.
\item Manipulate environment - Running the program itself and altering outside parameters (e.g., network state, files on disc, etc.) while observing how these affect the program. 
\item Monitor dynamic behavior - Run the program with additional tooling (e.g., packet capture, file monitoring) to see how it interacts with its environment.
\item Edit, recompile, and run - When an \RE{} changes the source code, then runs it to see what happens.
\item Fuzzing - Providing a series of varied inputs to the program and observing their effect on program behavior. These inputs can be selected manually or using automation.
\item Compare to known implementation - Running a known implementation of an algorithm used by the program (such as an encryption algorithm) and comparing the known implementation's results to the program's results.
\end{itemize}

\subsubsection{Static Analysis}
\begin{itemize}
\item Read code line-by-line - The \RE{} simply processes the state of the program in their head by running through the code line-by-line.
\item Scan beacons - Scanning through the code quickly and without much detail in the interest of identifying important beacons.
\item List function imports/strings - Listing out the functions imported or the strings used in the program.
\item Search for specific string - The \RE{} checks to see if a specific string is used.  This is performed either manually (i.e., scrolling through and scanning the code) or with the help of a search tool.
\item List file metadata - Listing out the metadata of a specific file such as its size or type.
\item Review differences from prior releases - The \RE{} looks at how the program has changed from version to version.
\item Reconstruct a data structure - The \RE{} writes out a variable's data structure in psuedo-code by making inferences from the binary.
\item Control flow analysis - Analyzing some path through the code on specific control flow inputs.
\item Data flow analysis - Analyzing some path through the code by data as it is passed between variables and through memory.
\item Symbolic execution - The \RE{} determines the set of symbols and expressions representing possible values of data at a specific point in the program.
\item Function call cross-referencing - The \RE{} determines where a particular function is called in the code.
\item Compare to known implementation - The \RE{} compares the code they believe is performing a particular algorithm or function to code from an outside source that is known to perform that algorithm or function.
\item Reimplementation - The \RE{} writes a program to perform the behaviors they believe the program under investigation is performing.  They then compare their implementation to the program under inspection to determine whether they are the same or identify differences.
\end{itemize}

\subsubsection{Metadata Review}
\begin{itemize}
\item check security mitigations - Checking the security restrictions or mitigations. 
\end{itemize}

\subsubsection{Method Interactions}
\begin{itemize}
\item View static and dynamic representation together - When the \RE{} views both static and dynamic code representations on their screen at the same time. For example, if they have a disassembler open reading code line-by-line side-by-side with a debugger.
\item Static to Dynamic - When the \RE{} first uses a static method, then uses information from this to inform the use of a dynamic method.
\item Dynamic to Static - When the \RE{} first uses a dynamic method, then uses information from this to inform the use of a static method.
\item Combined - When the \RE{} uses dynamic and static methods in concert, constantly passing information back and forth between static and dynamic methods.
\end{itemize}

\subsection{Decision}
For each decision point, we coded both the reasoning behind the \RE{}'s decision (\emph{reason}) and type of decision the \RE{} made (\emph{type}).

\subsubsection{Reason}

\begin{itemize} 
\item State of investigation - The \RE{} makes a decision based on where they are within the investigation process. For example, the \RE{} may choose to list APIs called because that is always their first step when reverse engineering a new program.
\item Strategy - The decision is dictated by an overarching \RE{} strategy. For example, the \RE{} may choose to look at functions in descending order according to their size.
\item Function prototype - The \RE{} made a decision based on a function's prototype (input and output types, number of arguments, etc). For example, if a function is passed a large number of function pointers as arguments, then it might be starting many threads and would be interesting to investigate.
\item Specific sub-goal - The decision was made because it was necessary to complete some other task.
\item Control flow path - The decision was dictated by the control flow path that was currently being followed.
\item Data flow path - The decision was dictated by the data flow path that was currently being followed.
\item Proximity to interesting information - The decision about some element was made because it is physically nearby some interesting information in the code.
\item Prior experience - The \RE{} made their decision based on prior reverse engineering experience.
\item Program metadata - The \RE{} made their decision based only on program metadata.
\item Observed behavior - The decision was made based on an understanding of what the program was actually doing.
\end{itemize} 

\subsubsection{Type}

\begin{itemize} 
\item Function/code to analyze - Which code segments or functions to attempt to analyze.
\item Analysis inputs - Which inputs to use when performing a simulation method.
\item Order of functions/code to analyze - The order in which the reverse engineer analyzes portions of code or functions.
\item Simulation method to use - Which simulation method to employ at that particular moment.
\end{itemize}

\end{document}